\documentclass[a4paper,11pt]{article}
\usepackage{pos}

\usepackage{physics}

\newcommand{\NN}{\nonumber\\ &}
\usepackage{graphicx}
\usepackage{amsmath}
\usepackage{amsfonts}
\usepackage{float}
\usepackage{hyperref}

\title{The Light Quark Connected Hadronic Vacuum Polarization Contribution to the muon anomaly via Sparsened Meson Fields}
 \ShortTitle{HVP$_{lqc}$ contribution to muon $g - 2$ via Sparsened Meson Fields}

\author*[a]{Vaishakhi Moningi}
\author[b]{Christopher Aubin}
\author[a]{Thomas Blum}
\author[c,d]{Maarten Golterman}
\author[a]{Luchang Jin}
\author[d]{Santiago Peris}

\affiliation[a]{Dept. of Physics, Univ. of Connecticut, \\
Storrs, CT 06269, USA}

\affiliation[b]{Dept. of Physics \& Engineering Physics, Fordham Univ.,\\
Bronx, NY 10458, USA}

\affiliation[c]{Dept. of Physics and Astronomy, San Francisco State Univ., \\San Francisco, CA 94132, USA}

\affiliation[d]{Dept. of Physics and IFAE-BIST, Univ. Autònoma de Barcelona, \\
E-08193 Bellaterra, Barcelona, Spain}

\emailAdd{vaishakhi.moningi@uconn.edu}

\abstract{We present an update on our determination of the light-quark connected contribution to the hadronic vacuum polarization (HVP) of the muon anomalous magnetic moment, $a_\mu$, on a finer lattice with  2+1+1 highly-improved staggered quark (HISQ) ensemble from the MILC collaboration with physical pion mass, 0.042 fm lattice spacing, and size $144^3 \times 288$ sites. Within the low-mode averaging (LMA) framework, the HVP correlator is decomposed into low-low (LL), high-low (HL), low-high (LH) and high-high (HH) components. Since the LL part dominates the total statistical uncertainty but is also the most computationally expensive to evaluate, we implement a sparsening strategy to construct the meson fields efficiently. This approach significantly reduces the computational cost while preserving signal quality. By combining the sparsened LL contribution with HL, LH and HH components, we achieve an improved determination of the light-quark connected HVP contribution to $a_\mu$.}

\FullConference{The 42nd International Symposium on Lattice Field Theory (LATTICE2025)\\
2-8 November 2025\\
Tata Institute of Fundamental Research, Mumbai, India\\}

\begin{document}
\maketitle

\section{Introduction}
The anomalous magnetic moment of the muon is defined as $a_\mu=(g-2)/2$, where $g$ is the {Land\'e} g-factor, proportional to the muon's intrinsic magnetic moment. In Dirac's relativistic quantum theory $g=2$ exactly, but in the Standard Model (SM) of particle physics it gets tiny corrections from the electromagnetic, strong, and weak interactions.  Currently it offers one of the most stringent tests of the SM because it has been measured and computed with high precision.
The new experimental world average, dominated by the measurements at Fermi National Accelerator Laboratory (FNAL), is $a_\mu = 1165920715(145) \times 10^{-12}$ (124 ppb). The measurements at FNAL have improved the precision on the world average by over a factor of 4 \cite{collaboration2025measurementpositivemuonanomalous}. A commensurate theoretical effort is necessary to match the expected experimental precision.

\section{Theoretical framework}
The leading-order hadronic vacuum polarization contribution to the muon anomalous magnetic moment $a_\mu $ is obtained via~\cite{Lautrup:1971jf,deRafael:1993za,Blum:2002ii}
\begin{eqnarray}
\label{eq:amu}
    a_\mu^{\rm HVP} &=& 2 \int_{0}^{\infty}dt ~ ~w(t') C(t')
\end{eqnarray}
where the kernel $w(t)$
comes from quantum electrodynamics (QED) \cite{Blum:2002ii,Bernecker:2011gh}
and $C(t)$ is the Euclidean time correlation function of two electromagnetic currents, summed over all quark flavors and averaged over spatial directions to project onto zero spatial momentum. 

The aim of lattice calculations is to efficiently obtain $C(t)$ with as much precision as possible. To do this we use the techniques of low-mode averaging (LMA)~\cite{Giusti:2004yp,DeGrand:2004qw} and all-mode averaging (AMA)~\cite{Blum:2013xva}. Both methods rely on the spectral decomposition of the quark propagator in terms of the eigenvectors of the lattice Dirac operator. The quark propagator from source point $y$ to sink point $x$, $S(x,y)$ mathematically is given by the inverse of the Dirac operator,
\begin{eqnarray}
S(x,y)= D^{-1}(x,y) =\sum_{\lambda\leq\lambda_N}\frac{\langle x|\lambda\rangle\langle\lambda|y\rangle}{\lambda}
+\sum_{\lambda>\lambda_N}\frac{\langle x|\lambda\rangle\langle\lambda|y\rangle}{\lambda}=
S_L+S_H,
\label{eq:spectral}
\end{eqnarray}
where the RHS shows the spectral decomposition divided into two pieces, one for the low modes up to {$\lambda_N$} and one for the rest, or high modes. Accordingly, we separate $C(t)$ into four parts: low-low (LL), low-high (LH), high-low (HL), and high-high (HH), 
\begin{equation}
    C(t)= {\frac{1}{3}}\sum_{\vec x, \vec y\textcolor{blue}{,i}} {\rm Tr}\,\gamma_i S(x,y)\gamma_i S(y,x)= C_{LL}+C_{LH}+C_{HL}+C_{HH},
    \label{C(t)_sep}
\end{equation}
In practice, once the low modes are determined, $S_H$ is determined by computing
the inverse of the deflated Dirac operator, using the conjugate gradient algorithm. So, not only do the low modes significantly enhance our statistics through LMA, but they also dramatically accelerate the computation of the high mode part.

\section{Lattice details}
The lattice computation presented here extends the simulations reported in Ref.~\cite{Aubin:2022hgm}. In the present work, we extend the analysis by incorporating a new fine ensemble, 144c, on which we compute the two-point correlation functions that together reconstruct the full correlator of Eq.~(\ref{C(t)_sep}). The results from the remaining five ensembles are taken from our previous calculation. 
All calculations are performed using gauge-field configurations generated by the MILC Collaboration with 2+1+1 flavors of HISQ quarks \footnote{Our 48II
ensemble in Table \ref{tab:lattice_data} is a CalLat ensemble \cite{hall2025}}. On the 144c ensemble, we compute 8000 (4000) low modes of the full (preconditioned) Dirac operator, enabling improved control of the long-distance behavior of the correlator. The lattice spacings are set using using the ${w_0}= 0.17187(68)$ fm which is  is determined using the mass of $\Omega^-$ baryon \cite{bazavov2025high} and {$w_0/a$} are obtained from v1 of Ref.~\cite{bazavov2025hadronicvacuumpolarizationmuon}

\begin{table}[htbp]
    \centering
    \resizebox{0.6\textwidth}{!}{
    \begin{tabular}{|l|l|l|l|l|l|}
    \hline
    Label & $m_\pi$ (MeV) & $a$ (fm) & ~~~~$N_S^3\times N_T$ & $N_{\text{low}}$ & \begin{tabular}[c]{@{}l@{}}\# configs\\ (LL-HL-HH)\end{tabular} \\ \hline
    144c & 134.5 & 0.04262 & $144^3\times 288$  & 4000 & 20-22-27    \\ \hline
    \end{tabular}%
    }
    \caption{ Lattice simulation details. “$N_{\text{low}}$” is the number of low-modes of the preconditioned Dirac operator. The number of configurations used for measurements in this study are given in the last column. }
    \label{tab:config}
\end{table}

\section{Motivation}
Obtaining our target precision becomes increasingly demanding on larger physical volumes, particularly for the 144c ensemble. We implement the separation of $C(t)$ as mentioned above in Eq.~(\ref{C(t)_sep}) in {Ref.}~\cite{Moningi:2025dlf}. This decomposition avoids redundant Dirac solves that would otherwise arise when the HH and HL terms are computed together and the LL contribution must be subtracted separately. Using this framework, the HL contribution is evaluated by forming stochastic linear combinations of low-mode sources on each time slice, reducing the number of solves to $N_T \times N_{\mathrm{hits}}$, where $N_T$ is the number of time slices. Here, $N_{\mathrm{hits}}$
 denotes the number of independent stochastic samples per time slice. For each hit, a new set of random numbers is combined with the low-mode eigenvectors to form a stochastic source, and multiple hits are averaged to reduce the stochastic error in the HL estimate.
This yields an improvement of $27\%$ in the long-distance region ($2.4$–$3.2$ fm) in $64$c ensemble for 10 hits.
On larger lattices like 144c, only a single stochastic hit for HL was feasible due to the high computational cost.

Nevertheless, the LL term remains both the most computationally expensive and the dominant source of statistical uncertainty in the long-distance tail in Fig.~(\ref{fig:amu integrand}). This motivates us to further optimize the LL contribution, which is the focus of the present work.

\section{Low-low contribution to $C(t)$}
On the lattice the local electromagnetic current is not conserved, so we use the point-split conserved current
\begin{align}
    J^\mu(x) &= -\frac{1}{2}\eta_\mu(x)(\Bar{\chi}(x+\hat{\mu})U^\dagger_\mu(x)\chi(x)+\Bar{\chi}(x)U_\mu(x)\chi(x+\hat{\mu}))
    \label{eq:current}
\end{align}
where $U_\mu$ are the gauge links for gauge invariance, $\chi(x)$ are the single spin component staggered fermion fields, and $\eta(x)$ arise from the spin diagonalization of the fermion action. 
To efficiently isolate the low-mode contribution, it is convenient to introduce the meson field, defined in the eigenmode basis of the Dirac operator as $
        \left(\Lambda_\mu(t)\right)_{n, m}=\sum_{\vec x}\expval{n|x} U_\mu(x)\expval{ x+\mu | m}(-1)^{(m+n)(x+m)}$,
where $n,m$ label eigenmodes of the Dirac operator, and the phases have been absorbed into the link variables, $U_\mu(x)\eta_\mu(x)\to U_\mu(x)$. Expressed in terms of these meson fields, the low–low (LL) contribution to the correlator takes a compact form,
\begin{align}
    C_{LL}= \frac{1}{4} \sum_{m, n} 
    \sum_{\vec{x}} \frac{1}{\lambda_m\lambda_n}&[ \Lambda^\dagger_\mu(x)_{mn}\Lambda^\dagger_\nu(y)_{nm} + \Lambda^\dagger_\mu(x)_{mn}\Lambda_\nu(y)_{nm} \NN+\Lambda_\mu(x)_{mn}\Lambda^\dagger_\nu(y)_{nm} + \Lambda_\mu(x)_{mn}\Lambda_\nu(y)_{nm}],
\end{align}
where $\lambda_n$ is shorthand for $i\lambda_n+m$.

\subsection{Sparsening of Meson Fields}
The cost of the meson field scales linearly in the size of the lattice and quadratically with the number of eigenvectors which is prohibitive when both are large as in the case of the $144^3\times288$ lattice. To significantly speedup the calculation, we “sparsen” the eigenvectors by omitting some number of sites in a regular pattern, $i.e.$, omit $s$ consecutive points, evenly spaced, in each direction. 
This is increasingly effective as $a\to 0$ since nearby points will be more and more correlated, and including them in the average does not meaningfully improve the statistical error. Sparsening also drastically reduces the memory footprint which is important if later we decide to increase the number of eigenvectors. 

To ensure we preserve the spin-taste structure of the staggered fermion currents, the sparsening is done by choosing the location for a {\it hypercube} randomly on a given time-slice, and then we omit every $s$ number of hypercubes in each spatial direction. In other words we always keep all points in a kept hypercube. By randomly choosing the initial hypercube on a time-slice we are guaranteed to project onto zero spatial momentum in the average. Sparsening by a factor $(s,t)$ reduces the size of eigenvectors required to compute our meson fields from $N_{\rm{S}}^3 \times N_{\rm{T}}$ to $(N_{\rm{S}}/s)^3 \times (N_{\rm{T}}/t)$. This reduction lead to a substantial increase in computational speed for the meson fields. Importantly, sparsening is limited to a factor of 4 in each dimension to avoid a noticeable increase in noise.

\begin{figure}[h]
    \centering
    \includegraphics[width=0.47\textwidth]{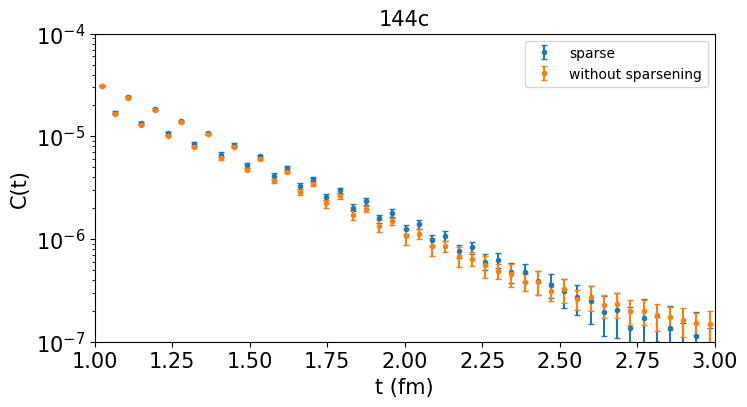}
    \includegraphics[width=0.48\textwidth]{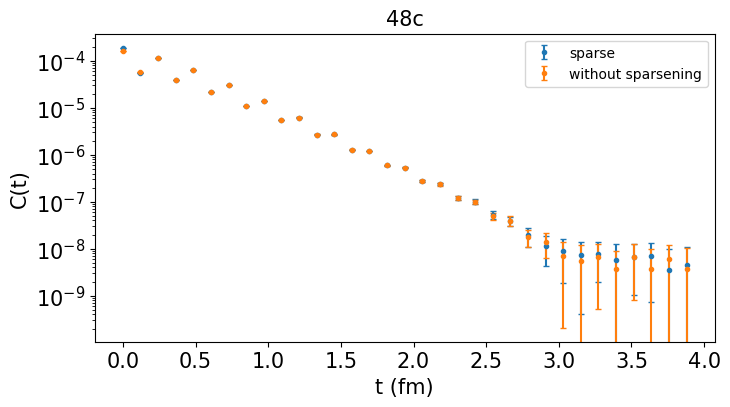}
    \caption{Left: Sparsening of factor (s=4,t=1) on 6 different configurations on $144$c ensemble. Right: Sparsening of factor (s=2,t=1) on the same 24 configurations on $48$I ensemble }
\end{figure}


\section{Results}
\vspace{-7pt}
\begin{figure}[h]
    \begin{center}
        \includegraphics[width=0.343\textwidth]{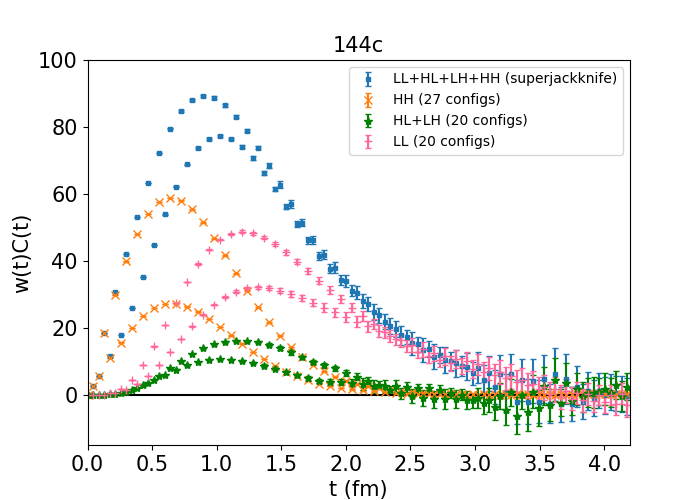}\hskip .0\textwidth
        \includegraphics[width=0.327\textwidth]{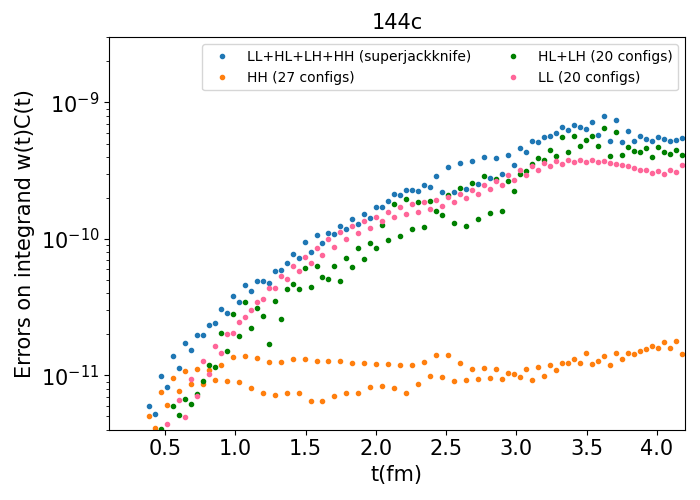}
        \includegraphics[width=0.32\textwidth]{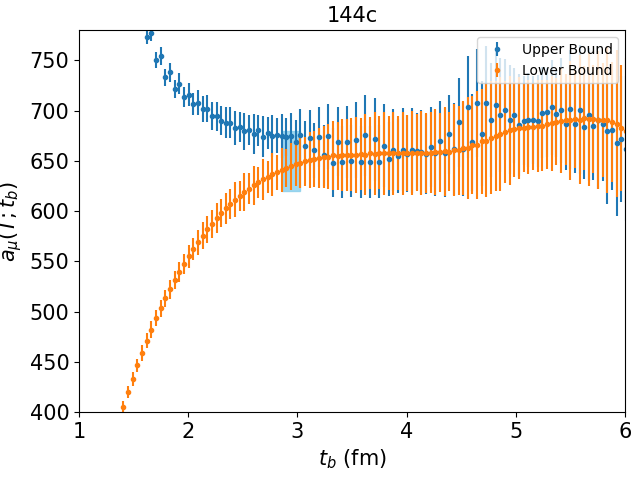}
        \caption{Left: Summand in $w(t)C(t)$ Eq.~(\ref{C(t)_sep}). Middle: Quantitative breakdown of the statistical error in terms of LL, HH, HL on the summand. Right: $(a_\mu(T; t_b))$ using the bounding method, as a function of the switch-point $t_b$.}
        \label{fig:amu integrand}
    \end{center}
\end{figure}
Lattice simulations on the 144c ensemble yield the correlator $C(t)$ defined in Eq.~(\ref{eq:amu}). We then define the trapezoid integration on the lattice,
\begin{align}
  a_\mu(t) \xrightarrow{a\to 0}  2\int_0^{t/2} dt'w(t')C(t') = 2 a\sum_{n=0}^{[t/2a]-1} w(na)C(na)   + aw([t/2a]a)C([t/2a]a),
\end{align} 
where the maximum value of  $t/a$ is $ N_T$. 
The bounding method \cite{Borsanyi:2017zdw,Blum:2018mom} on the correlation function is also applied in the time integration. We define the upper and lower bound of the two-point correlation function as, for $t\le T,~ C_b(t) = C(t)$, and for $t > T,~C_b(t) = 0$ (lower bound) and $C_b(t) = C(T)\,e^{-E_0(t-T)}$ (upper bound) where
$E_0 = {2\sqrt{m_\pi^2+(2\pi/L)^2}}$. $E_0$ denotes the {(non-interacting)} two-pion energy state. By using this method, we do not have to sum over the whole range of time slices to compute $a^{\text{HVP}}_\mu$ , which contain the noisy long-distance tail. Instead, an estimate can be got from the overlap between {the} two bounds in large $T$. The statistical errors on the averages are computed using the jackknife method. 

We incorporate finite-volume (FV), taste-breaking (TB) and pion-mass retuning effects into our lattice determination, see Table \ref{tab:lattice_data}. 
The analysis proceeds in stages. These corrections are computed in NLO and NNLO SChPT, as well as in the chiral vector model (ChVM) 
developed specifically for the case of staggered fermions in Ref.~\cite{chakraborty2017hadronic} and applied further in {Refs.~}\cite{davies2020hadronic,borsanyi2021leading,Aubin:2022hgm}. For each ensemble, we first apply the {NNLO} finite-volume correction, the corresponding taste-breaking correction, and the adjustment accounting for pion-mass mistuning. The corrected results are then extrapolated to the continuum limit using an $a^2$ ansatz based on six ensembles (five different lattice spacings). Our extrapolated values from these fits are as follows:
\begin{table}[h]
    \centering
    \resizebox{0.55\textwidth}{!}{%
    \begin{tabular}{|c|c|c|c|}
    \hline
    Ensemble & $a_\mu^{\mathrm{HVP, lqc}}$ & $a_\mu^{\mathrm{w1, lqc}}$ & $a_\mu^{\mathrm{w2, lqc}}$ \\ \hline
    144c & 658.4 (15.3) (4.7)  & 207.36 (49) (34) & 98.4 (1.9) (1.0) \\ \hline
    96c  & 609.3 (13.0) (4.4)  & 206.14 (38) (34) & 95.2 (2.4) (1.1) \\ \hline
    64c  & 599.8 (9.3) (4.3)  & 205.27(26)( 37)    & 88.9 (1.3) (1.1) \\ \hline
    48I  & 552.3(8.6) (4.0)  & 201.96 (53) (42) & 78.2 (1.4) (1.0) \\ \hline
    48II & 522.7 (8.0) (3.8) & 199.78 (42) (44) & 72.5 (1.2) (1.0) \\ \hline
    32c  & 505.3 (6.9) (3.7) & 200.64 (68) (45) & 71.3 (1.0) (1.0) \\ \hline
    \end{tabular}%
    }
    \caption{Lattice results for $a_\mu^{\mathrm{HVP, lqc}}$, the 0.4–1.0 fm window $a_\mu^{\mathrm{w1, lqc}}$
and the 1.5–1.9 fm window $a_\mu^{\mathrm{w2, lqc}}$. The first error is statistical,
the second error from scale setting.}
    \label{tab:lattice_data}
\end{table}

\begin{figure}[h]
    \centering
    \includegraphics[width=6.6cm]{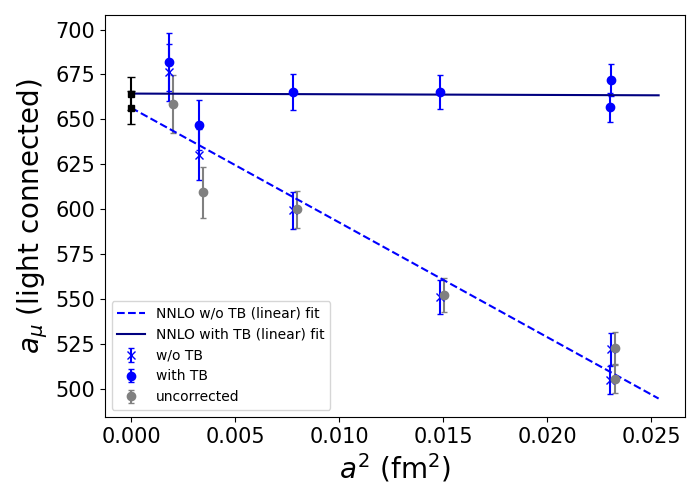}   
     \includegraphics[width=6.6cm]{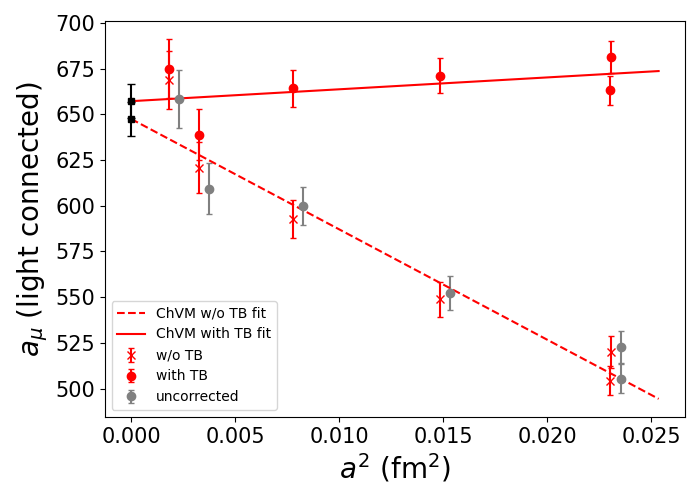} 
     \caption{Left: SChPT NNLO fits -- linear, blue (dashed=no TB, solid=TB). Right: ChVM fits -- linear, red (dashed=no TB, solid=TB). Data points match fit colors; continuum limits in black. Gray points show uncorrected values from Table \ref{tab:lattice_data}.}
    \label{fig:fullHVPfits}
\end{figure}

\vspace{-4pt}
\begin{align}
\text{NNLO}: a_\mu^{\mathrm{HVP,lqc}} &= (656.5 \pm 9.2)\times 10^{-10} \;\; \text{without TB},
\quad\; (664.4 \pm 9.2)\times 10^{-10} \;\; \text{with TB} \\
\text{ChVM}: a_\mu^{\mathrm{HVP,lqc}} &= (647.5 \pm 9.2)\times 10^{-10} \;\; \text{without TB},
\quad\; (657.3 \pm 9.2)\times 10^{-10} \;\; \text{with TB}
\end{align}

We take the average of these fits with and without taste breaking in their respective scheme as our best value
\begin{align}
   \text{NNLO}: a_\mu^{\mathrm{HVP,lqc}} &= (660.5\pm 9.2 \pm 4.0 \pm 1.9 \pm 0.3) \times 10^{-10}  = 661(10)\times 10^{-10}  \label{fullHVP-NNLO}
    \\ \text{ChVM}: a_\mu^{\mathrm{HVP,lqc}} &= (652.4\pm 9.2 \pm 4.9) \times 10^{-10}  = 652(10)\times 10^{-10}  \label{fullHVP-ChM}
\end{align}

For all fits, we have assumed the statistical errors to be uncorrelated, since the data are obtained on different ensembles and the scale-setting error is  folded into the fit error, as we
took the scale-setting errors of Table \ref{tab:lattice_data} into account in our fits, assuming them to be $100\%$ correlated.
  The four errors in Eq.~(\ref{fullHVP-NNLO}) are statistical error from the fit, half the difference between the two fits with and without TB, and the errors on the NNLO FV and retuning corrections closest to the average ($144$c ensemble). The error in the second equality in Eqs.~(\ref{fullHVP-NNLO}) and (\ref{fullHVP-ChM}) is obtained by adding these errors in quadrature.
  
    \begin{figure}[h]
        \centering
        \includegraphics[width=0.5\linewidth]{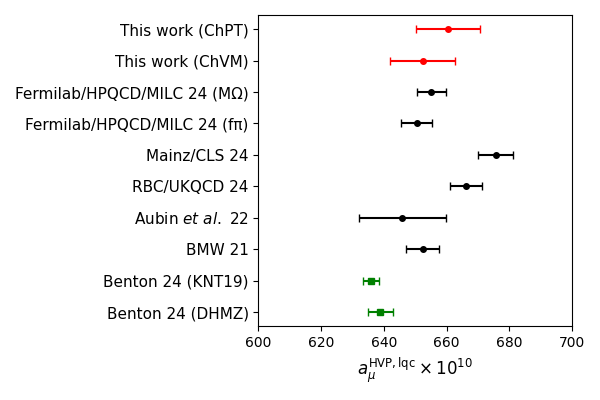}
        \caption{ Comparison of our lattice determinations
            for $a_\mu^{\mathrm{HVP,lqc}}$(red circles)  with other lattice results (black circles)
            and data-driven results (green
            squares) }
            \label{fullHVP-compare}
    \end{figure}

In Fig.~(\ref{fullHVP-compare}) we compare our result for $a_\mu^{\mathrm{HVP,lqc}}$ with previous lattice-QCD calculations as
well as the data-driven evaluations. With our updated results in Eq.~\ref{fullHVP-NNLO}(resp. (\ref{fullHVP-ChM})), we
find significances in the differences of $0.47\sigma~(0.25\sigma)$ with Fermilab/HPQCD/MILC 24 using $M_\Omega$ and $0.88\sigma~(0.16\sigma)$ using $f_\pi$ \cite{bazavov2025hadronic},
$0.50\sigma~(1.19\sigma)$ with RBC/UKQCD 24 \cite{blum2025long}, 
$1.31\sigma~(1.98\sigma)$ with Mainz/CLS 24 \cite{djukanovic2025hadronic}, $0.85\sigma~(0.37\sigma)$ with Aubin 22 \cite{Aubin:2022hgm} and $0.70\sigma~(0.0\sigma)$ with BMW 21 \cite{borsanyi2021leading} . Finally, we find that our result differs from the data-driven evaluation such as Benton~24 by
$2.35\sigma~(1.55\sigma)$ using KNT and 
$1.97\sigma~(1.21\sigma)$ using DHMZ \cite{benton2025data}. This tension clearly indicates  the need for more precise calculation.


\subsection{Window Quantities}
We further investigate $a_\mu$ for the intermediate window \cite{Blum:2018mom} as it is used to make more precise comparisons with others since it focuses on medium-time-distance region where the results are more precise. 
\begin{align}
    a_\mu = 2 \sum_{t=0}^{T/2} C(t) w(t) \left( \Theta(t, t_0, \Delta) - \Theta(t, t_1, \Delta) \right), ~~~ \text{with}~~ \Theta(t, t', \Delta)=\frac{1}{2}\left(1+\tanh\frac{t-t'}{\Delta}\right)
\end{align}
\subsubsection{Window 0.4-1.0 fm}
We show $a_\mu^{\mathrm{W1,lqc}}$ as a function of $a^2$. 
    \begin{figure}[h]
        \centering
        \includegraphics[width=6.6cm]{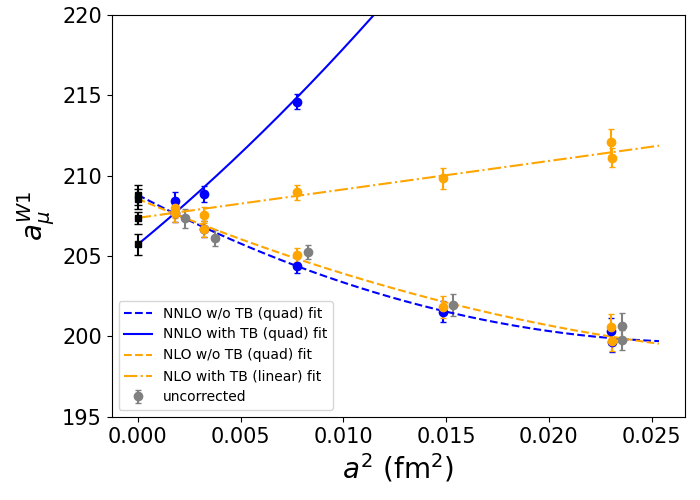}   
         \includegraphics[width=6.6cm]{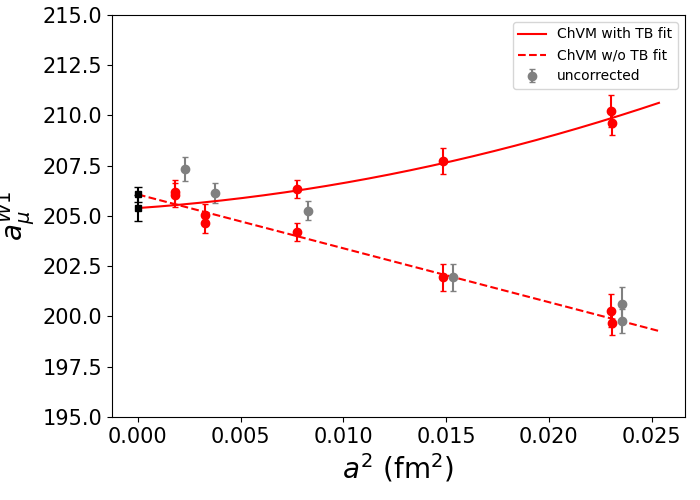}
         \caption{Left: SChPT fits - NNLO quadratic, blue (dashed = no TB, solid = with TB). NLO yellow: quadratic (dotted = no TB) and linear (dot-dashed = with TB). Right: ChVM quadratic, red (dashed = no TB, solid = with TB). Continuum limits in black. Gray points show uncorrected values from Table \ref{tab:lattice_data}.}
        \label{fig:fitsw1}
    \end{figure}
Our extrapolated values from these fits are as follows:
    \begin{align}
        \text{ChVM}:a_\mu^{\mathrm{W1,lqc}} &= 
                         (206.07 \pm 0.38) \times 10^{-10} ~~\text{without TB}, ~(205.39 \pm 0.64) \times 10^{-10} ~~\text{with TB }
    \end{align}

Applying chiral perturbation theory (ChPT) to estimate finite-volume and taste-breaking corrections for the intermediate window quantity $a_\mu^{\mathrm{W1}}(0.4,1.0;0.15)$ (with values of the arguments in fm) is not fully justified, since this window is restricted to distances below 1 fm and is therefore not dominated by purely long-distance physics where ChPT is expected to be reliable. This motivates the use of phenomenological models such as ChVM. However, while such models can describe the data within a limited range, they cannot provide controlled extrapolations beyond it. Consequently, for this window, continuum and infinite-volume limits cannot be obtained from lattice data alone without introducing model dependence, as no fully reliable EFT description is available in this regime.

For this reason, we also consider a longer-distance window, $a_\mu^{\mathrm{W2}}(1.5,1.9;0.15)$ (with values of the arguments in fm), which is more firmly in the long-distance regime and therefore better suited for treatment within ChPT.

\subsubsection{Window 1.5-1.9 fm}
    \begin{figure}[h]
        \centering
        \includegraphics[width=6.6cm]{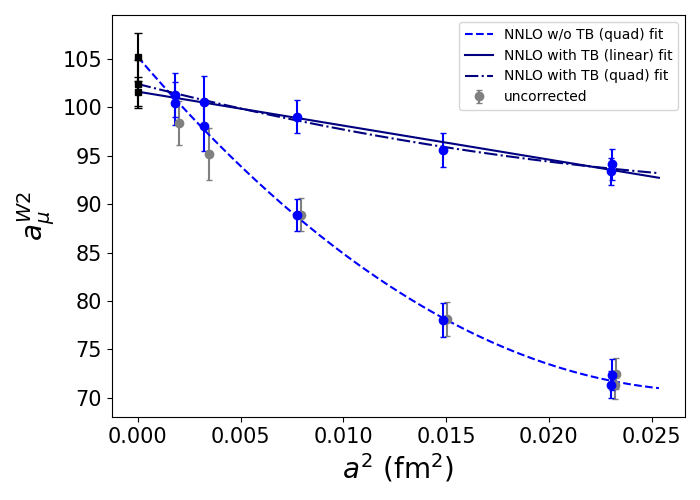}   
         \includegraphics[width=6.6cm]{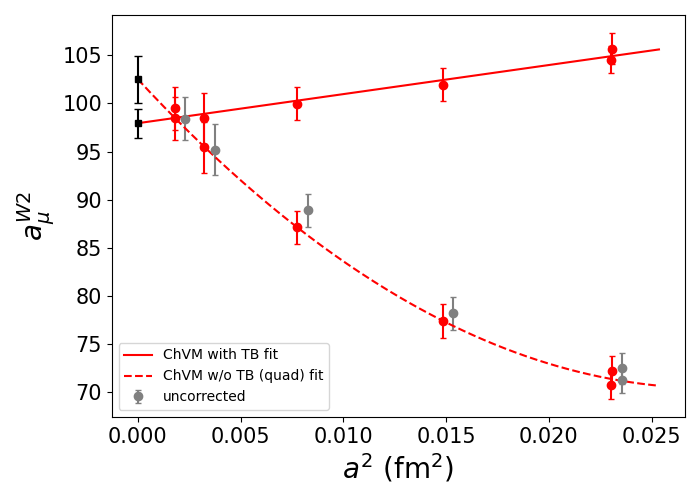} 
         \caption{
         Left: SChPT NNLO fits - quadratic, blue (dashed=no TB, dotted-dashed=TB) and linear, blue (solid=TB). Right: ChVM fits - quadratic, red (dashed=no TB, solid=TB). Data points match fit colors; continuum limits in black. Gray points show uncorrected values from Table \ref{tab:lattice_data}.}
        \label{fig:fitsw2}
    \end{figure}
    \vspace{-20pt}
    \begin{align}
       \text{NNLO}:a_\mu^{\mathrm{W2,lqc}} &= (105.2 \pm 2.5) \times 10^{-10} ~~\text{without TB},~
                         (101.6 \pm 1.5) \times 10^{-10} ~\text{with TB linear} \\
\text{ChVM}:  a_\mu^{\mathrm{W2,lqc}} &=(102.5 \pm 2.5) \times 10^{-10} ~~\text{without TB},~(97.9 \pm 1.5) \times 10^{-10} ~~\text{with TB}
    \end{align}

\section{Conclusion}
In our earlier analysis, based on ensembles up to 96c, we obtained $a_\mu^{\mathrm{HVP,lqc}} = 646(14)\times 10^{-10}$ using NNLO SChPT and $638(14)\times 10^{-10}$ using ChVM. With the addition of the 144c ensemble, these values become $661(10)\times 10^{-10}$ and $652(10)\times 10^{-10}$, respectively, corresponding to a reduction in the total uncertainty by a factor of $1.4$. 

Compared to our previous work, the present calculation achieves a quantitative improvement in precision primarily through improved control of the light-quark connected contribution, driven by the 144c ensemble. Although our total uncertainty remains larger than the most precise determinations, further reduction is expected by increasing statistics on the fine lattices; in particular, we plan to double up the statistics to reduce the error and achieve our sub-precent goal.

In Fig.~(\ref{fig:amu integrand} (middle)), we observe that the HL contribution, while subdominant in the total integral, develops a rapidly growing uncertainty at large Euclidean times. One straightforward way to control this behavior is to increase the number of hits per configuration, though this becomes computationally expensive on 144c. An alternative and potentially more efficient strategy is to develop a multi-grid \cite{brower:2018} or AMA-inspired machine learning framework \cite{Blum:2025woh} that predicts the long-distance HL contribution at significantly reduced cost. Such an approach could provide improved statistical precision in the tail region without a proportional increase in computational resources.
\begin{table}[H]
    \centering
    \resizebox{0.6\textwidth}{!}{%
    \begin{tabular}{|l|l|l|l|}
    \hline
    a (fm) & $a_\mu^{\mathrm{HVP,lqc}}\times 10^{-10}$ & $a_\mu^{\mathrm{W1,lqc}}\times 10^{-10}$ & $a_\mu^{\mathrm{W2,lqc}}\times 10^{-10}$ \\ \hline
    0.04262(17)   & 658(15)(5){[}15{]} & 207.3(5)(3){[}6{]}     & 98.4(1.9)(1.0){[}2.1{]}  \\ \hline
    0 (NNLO ChPT) & 661(9)(4){[}10{]}    &  & 103.4(2.5)(1.8){[}3.1{]} \\ \hline
    0 (ChVM)      & 652(9)(5){[}10{]}    & 205.7(6)(3){[}7{]}  & 100.2(2.5)(2.3){[}3.3{]}  \\ \hline
    \end{tabular}%
    }
    \caption{HVP contributions to the muon anomaly using NNLO SChPT and ChVM. First row errors: (statistical)(scale)[tot]; Second+Third row errors: (fit error)(sys)[tot]. Systematic error is taken as half the difference between the continuum extrapolated results with and without TB.}
\label{tab:summary}
\end{table}



\acknowledgments
We gratefully acknowledge support by the U.S.\ Department of Energy,
Office of Science, Office of High Energy Physics, under Awards DE-SC0013682 (MG), DE-SC0021147, DE-SC0010339 and DE-SC0026314 (LJ) and DE-SC0010339 (TB,VM). SP is supported by the Spanish Ministerio de Ciencia e Innovacion,
grants PID2020-112965GB-I00 and PID2023-146142NB-I00,
 and by the Departament de Recerca i Universities from Generalitat de Catalunya to the Grup de Recerca 00649 (Codi: 2021 SGR 00649). IFAE is partially funded by the CERCA program of the Generalitat de Catalunya. The authors acknowledge the \href{ http://www.tacc.utexas.edu}{Texas Advanced Computing Center} (TACC) at The University of Texas at Austin for providing computational resources on Frontera, Stampede 2, and Stampede 3 that have contributed to the research results reported in this proceedings.


{\small
\bibliographystyle{JHEP}
\bibliography{ref.bib}

@article{Blum:2002ii,
      author         = "Blum, Thomas",
      title          = "{Lattice calculation of the lowest order hadronic
                        contribution to the muon anomalous magnetic moment}",
      journal        = "Phys.Rev.Lett.",
      volume         = "91",
      pages          = "052001",
      doi            = "10.1103/PhysRevLett.91.052001",
      year           = "2003",
      eprint         = "/0212018",
      archivePrefix  = "arXiv",
      primaryClass   = "",
      reportNumber   = "RBRC-296",
      SLACcitation   = "%%CITATION = /0212018;%%",
}

@article{Bernecker:2011gh,
      author         = "Bernecker, David and Meyer, Harvey B.",
      title          = "{Vector Correlators in Lattice QCD: Methods and
                        applications}",
      journal        = "Eur.Phys.J.",
      volume         = "A47",
      pages          = "148",
      doi            = "10.1140/epja/i2011-11148-6",
      year           = "2011",
      eprint         = "1107.4388",
      archivePrefix  = "arXiv",
      primaryClass   = "",
      SLACcitation   = "%%CITATION = ARXIV:1107.4388;%%",
}

@article{Blum:2013xva,
      author         = "Blum, Thomas and Denig, Achim and Logashenko, Ivan and de
                        Rafael, Eduardo and Lee Roberts, B. and others",
      title          = "{The Muon (g-2) Theory Value: Present and Future}",
      year           = "2013",
      eprint         = "1311.2198",
      archivePrefix  = "arXiv",
      primaryClass   = "hep-ph",
      SLACcitation   = "%%CITATION = ARXIV:1311.2198;%%",
}

@article{Blum:2018mom,
      author         = "Blum, T. and Boyle, P. A. and Gulpers, V. and Izubuchi,
                        T. and Jin, L. and Jung, C. and Juettner, A. and Lehner,
                        C. and Portelli, A. and Tsang, J. T.",
      title          = "{Calculation of the hadronic vacuum polarization
                        contribution to the muon anomalous magnetic moment}",
      collaboration  = "RBC, UKQCD",
      journal        = "Phys. Rev. Lett.",
      volume         = "121",
      number         = "2",
      pages          = "022003",
      doi            = "10.1103/PhysRevLett.121.022003",
      year           = "2018",
      eprint         = "1801.07224",
      archivePrefix  = "arXiv",
      primaryClass   = "",
      SLACcitation   = "%%CITATION = ARXIV:1801.07224;%%"
}

@article{Borsanyi:2017zdw,
      author         = "Borsanyi, Sz. and others",
      title          = "{Hadronic vacuum polarization contribution to the
                        anomalous magnetic moments of leptons from first
                        principles}",
      collaboration  = "Budapest-Marseille-Wuppertal",
      journal        = "Phys. Rev. Lett.",
      volume         = "121",
      year           = "2018",
      number         = "2",
      pages          = "022002",
      doi            = "10.1103/PhysRevLett.121.022002",
      eprint         = "1711.04980",
      archivePrefix  = "arXiv",
      primaryClass   = "",
      SLACcitation   = "%%CITATION = ARXIV:1711.04980;%%"
}

@article{Giusti:2004yp,
      author         = "Giusti, Leonardo and Hernandez, P. and Laine, M. and
                        Weisz, P. and Wittig, H.",
      title          = "{Low-energy couplings of QCD from current correlators
                        near the chiral limit}",
      journal        = "JHEP",
      volume         = "04",
      year           = "2004",
      pages          = "013",
      doi            = "10.1088/1126-6708/2004/04/013",
      eprint         = "/0402002",
      archivePrefix  = "arXiv",
      primaryClass   = "",
      reportNumber   = "BI-TP-2004-03, CPT-2003-P-4622, DESY-04-009,
                        FTUV-04-0203, IFIC-04-04, MPP-2004-6",
      SLACcitation   = "%%CITATION = /0402002;%%"
}

@article{DeGrand:2004qw,
      author         = "DeGrand, Thomas A. and Schaefer, Stefan",
      title          = "{Improving meson two point functions in lattice QCD}",
      journal        = "Comput. Phys. Commun.",
      volume         = "159",
      year           = "2004",
      pages          = "185-191",
      doi            = "10.1016/j.cpc.2004.02.006",
      eprint         = "/0401011",
      archivePrefix  = "arXiv",
      primaryClass   = "",
      reportNumber   = "COLO-HEP-497",
      SLACcitation   = "%%CITATION = /0401011;%%"
}

@article{Lautrup:1971jf,
      author         = "Lautrup, B. e. and Peterman, A. and de Rafael, E.",
      title          = "{Recent developments in the comparison between theory and
                        experiments in quantum electrodynamics}",
      journal        = "Phys. Rept.",
      volume         = "3",
      year           = "1972",
      pages          = "193-259",
      doi            = "10.1016/0370-1573(72)90011-7",
      SLACcitation   = "%%CITATION = PRPLC,3,193;%%"
}

@article{deRafael:1993za,
      author         = "de Rafael, Eduardo",
      title          = "{Hadronic contributions to the muon g-2 and low-energy
                        QCD}",
      journal        = "Phys. Lett.",
      volume         = "B322",
      year           = "1994",
      pages          = "239-246",
      doi            = "10.1016/0370-2693(94)91114-2",
      eprint         = "hep-ph/9311316",
      archivePrefix  = "arXiv",
      primaryClass   = "hep-ph",
      reportNumber   = "CPT-93-P-2962",
      SLACcitation   = "%%CITATION = HEP-PH/9311316;%%"
}

@article{Aubin:2022hgm,
    author = "Aubin, Christopher and Blum, Thomas and Golterman, Maarten and Peris, Santiago",
    title = "{Muon anomalous magnetic moment with staggered fermions: Is the lattice spacing small enough?}",
    eprint = "2204.12256",
    archivePrefix = "arXiv",
    primaryClass = "hep-lat",
    doi = "10.1103/PhysRevD.106.054503",
    journal = "Phys. Rev. D",
    volume = "106",
    number = "5",
    pages = "054503",
    year = "2022"
}

@article{Moningi:2025dlf,
    author = "Moningi, Vaishakhi and Aubin, Christopher and Blum, Thomas and Golterman, Maarten and Jin, Luchang and Peris, Santiago",
    title = "{Progress on computing the hadronic vacuum polarization contribution to the muon anomalous magnetic moment with staggered fermions}",
    doi = "10.22323/1.466.0247",
    journal = "PoS",
    volume = "LATTICE2024",
    pages = "247",
    year = "2025"
}

@article{borsanyi2021leading,
  title={Leading hadronic contribution to the muon magnetic moment from lattice QCD},
  author={Borsanyi, Sz and Fodor, Z and Guenther, JN and Hoelbling, C and Katz, SD and Lellouch, L and Lippert, T and Miura, K and Parato, L and Szabo, KK and others},
  journal={Nature},
  volume={593},
  number={7857},
  pages={51--55},
  year={2021},
  publisher={Nature Publishing Group UK London},
  doi={10.1038/s41586-021-03418-1},
  url={https://doi.org/10.1038/s41586-021-03418-1}
}

@article{collaboration2025measurementpositivemuonanomalous,
      title={Measurement of the Positive Muon Anomalous Magnetic Moment to 127 ppb}, 
      author={Donati, S and others},
      year={2025},
       journal = {Phys. Rev. Lett.},
      volume = {135},
      issue = {10},
      pages = {101802},
      numpages = {13},
      year = {2025},
      month = {Sep},
      publisher = {American Physical Society},
      doi = {10.1103/7clf-sm2v},
      url = {https://link.aps.org/doi/10.1103/7clf-sm2v} 
}

@article{bazavov2025hadronic,
  title = {Hadronic Vacuum Polarization for the Muon $g\ensuremath{-}2$ from Lattice QCD: Long-Distance and Full Light-Quark Connected Contribution},
   author={Bazavov, Alexei and Bernard, Claude W and Clarke, David A and Davies, Christine and DeTar, Carleton and El-Khadra, Aida X and G{\'a}miz, Elvira and Gottlieb, Steven and Grebe, Anthony V and Hostetler, Leon and others},
  collaboration = {Fermilab Lattice, HPQCD, and MILC Collaborations},
  journal = {Phys. Rev. Lett.},
  volume = {135},
  issue = {1},
  pages = {011901},
  numpages = {10},
  year = {2025},
  month = {Jul},
  publisher = {American Physical Society},
  doi = {10.1103/d583-yhfs},
  url = {https://link.aps.org/doi/10.1103/d583-yhfs}
}

@article{benton2025data,
  title={Data-driven results for light-quark connected and strange-plus-disconnected hadronic g- 2 short-and long-distance windows},
  author={Benton, Genessa and Boito, Diogo and Golterman, Maarten and Keshavarzi, Alexander and Maltman, Kim and Peris, Santiago},
  journal = {Phys. Rev. D},
  volume = {111},
  issue = {3},
  pages = {034018},
  numpages = {18},
  year = {2025},
  month = {Feb},
  publisher = {American Physical Society},
  doi = {10.1103/PhysRevD.111.034018},
  url = {https://link.aps.org/doi/10.1103/PhysRevD.111.034018}
}

@article{djukanovic2025hadronic,
  title={The hadronic vacuum polarization contribution to the muon g- 2 at long distances},
  author={Djukanovic, Dalibor and von Hippel, Georg and Kuberski, Simon and Meyer, Harvey B and Miller, Nolan and Ottnad, Konstantin and Parrino, Julian and Risch, Andreas and Wittig, Hartmut},
  journal={Journal of High Energy Physics},
  volume={2025},
  number={4},
  pages={1--57},
  year={2025},
  publisher={Springer},
  doi={10.1007/JHEP04(2025)098},
  url={https://doi.org/10.1007/JHEP04(2025)098}
  
}

@article{blum2025long,
  title={Long-Distance Window of the Hadronic Vacuum Polarization for the Muon g-2},
  author={Blum, T and Boyle, PA and Bruno, M and Chakraborty, B and Erben, F and G{\"u}lpers, V and Hackl, A and Hermansson-Truedsson, N and Hill, RC and Izubuchi, T and others},
  journal={Physical Review Letters},
  volume={134},
  number={20},
  pages={201901},
  year={2025},
  publisher={APS},
  doi = {10.1103/PhysRevLett.134.201901},
  url = {https://link.aps.org/doi/10.1103/PhysRevLett.134.201901}
}

@article{bazavov2025high,
  title={High-Precision Scale Setting with the Omega-Baryon Mass and Gradient Flow},
  author={Bazavov, Alexei and Bernard, Claude W and Clarke, David A and DeTar, Carleton and El-Khadra, Aida X and G{\'a}miz, Elvira and Gottlieb, Steven and Grebe, Anthony V and Heller, Urs M and Hostetler, Leon and others},
  journal = {Phys. Rev. D},
  pages = {--},
  year = {2026},
  month = {Jan},
  publisher = {American Physical Society},
  doi = {10.1103/t18c-bpqy},
  url = {https://link.aps.org/doi/10.1103/t18c-bpqy}
}

@article{chakraborty2017hadronic,
  title={Hadronic vacuum polarization contribution to a $\mu$ from full lattice QCD},
  author={Chakraborty, Bipasha and Davies, CTH and De Oliveira, PG and Koponen, J and Lepage, GP and (HPQCD Collaboration) and Van de Water, RS},
  journal={Physical Review D},
  volume={96},
  number={3},
  pages={034516},
  year={2017},
  doi={https://doi.org/10.1103/PhysRevD.96.034516},
  publisher={APS}
}

@article{davies2020hadronic,
  title={Hadronic-vacuum-polarization contribution to the muon’s anomalous magnetic moment from four-flavor lattice QCD},
  author={Davies, Christine TH and DeTar, C and El-Khadra, AX and Gamiz, E and Gottlieb, Steven and Hatton, D and Kronfeld, AS and Laiho, J and Lepage, GP and Liu, Yuzhi and others},
  journal = {Phys. Rev. D},
  volume = {101},
  issue = {3},
  pages = {034512},
  numpages = {19},
  year = {2020},
  month = {Feb},
  publisher = {American Physical Society},
  doi = {10.1103/PhysRevD.101.034512},
  url = {https://link.aps.org/doi/10.1103/PhysRevD.101.034512}
}

@article{bazavov2025hadronicvacuumpolarizationmuon,
      title={Hadronic vacuum polarization for the muon $g-2$ from lattice QCD: Complete short and intermediate windows}, 
      author={Bazavov, Alexei and Clarke, David A and Davies, Christine TH and DeTar, Carleton and El-Khadra, Aida X and G{\'a}miz, Elvira and Gottlieb, Steven and Grebe, Anthony V and Hostetler, Leon and Jay, William I and others},
  collaboration = {Fermilab Lattice, HPQCD, and MILC Collaborations},
  eprint={2411.09656},
      archivePrefix={arXiv},
      primaryClass={hep-lat},
      doi={https://doi.org/10.1103/PhysRevD.111.094508},
      url={https://arxiv.org/abs/2411.09656}, 

}

@article{brower:2018,
  title = {Multigrid algorithm for staggered lattice fermions},
  author = {Brower, Richard C. and Weinberg, Evan and Clark, M. A. and Strelchenko, Alexei},
  journal = {Phys. Rev. D},
  volume = {97},
  issue = {11},
  pages = {114513},
  numpages = {21},
  year = {2018},
  month = {Jun},
  publisher = {American Physical Society},
  doi = {10.1103/PhysRevD.97.114513},
  url = {https://link.aps.org/doi/10.1103/PhysRevD.97.114513}
}

@inproceedings{Blum:2025woh,
    author = "Blum, Thomas and Conigli, Alessandro and Geyer, Lukas and Kuberski, Simon and Segner, Alexander and Wittig, Hartmut",
    title = "{Machine-learning techniques as noise reduction strategies in lattice calculations of the muon $g-2$}",
    booktitle = "{41st International Symposium on Lattice Field Theory}",
    eprint = "2502.10237",
    archivePrefix = "arXiv",
    primaryClass = "hep-lat",
    reportNumber = "CERN-TH-2025-038, MITP-25-010",
    month = "2",
    year = "2025"
}

@article{hall2025,
      title={Signs of Non-Monotonic Finite-Volume Corrections to $g_A$}, 
      author={Zack B. Hall and Dimitra A. Pefkou and Aaron S. Meyer and Thomas R. Richardson and Raúl A. Briceño and M. A. Clark and Martin Hoferichter and Emanuele Mereghetti and Henry Monge-Camacho and Colin Morningstar and Amy Nicholson and Pavlos Vranas and André Walker-Loud},
      year={2025},
      eprint={2503.09891},
      archivePrefix={arXiv},
      primaryClass={hep-lat},
      url={https://arxiv.org/abs/2503.09891}, 
}
}

\end{document}